\documentclass[journal,twoside]{IEEEtran}

\usepackage[font=small,labelfont=normal]{caption}
\usepackage{amsthm}
\usepackage{fancyhdr}
\usepackage{kantlipsum}
\usepackage{graphicx}
\usepackage{amssymb}
\usepackage{amsmath}
\usepackage{cleveref}
\usepackage{cite}
\usepackage{stfloats}
\usepackage{color}
\usepackage{graphicx} 
\usepackage{bm}
\usepackage{amsmath}
\usepackage{booktabs}
\usepackage{mathtools}
\usepackage{amsmath}
\usepackage{multirow}
\usepackage{booktabs}
\usepackage{authblk}
\usepackage{dsfont}
\usepackage{epstopdf}
\usepackage{tcolorbox}
\usepackage{graphicx}
\usepackage[ruled]{algorithm2e}
\usepackage{pseudocode}
\DeclareMathAlphabet{\mathsf}{OML}{cmbr}{m}{it}

\Crefrangeformat{figure}{Fig. #3(#1)#4--#5(#2)#6}
\newtheoremstyle{mystyle}
  {0.5pt}
  {0.5pt}
  {\normalfont}
  {0.6pt}
  {\bfseries}
  {:}
  {.5em}
  {}
\theoremstyle{mystyle}

\newtheorem{remark*}{Remark}

\newtheorem{Discussion*}{Discussion}
\newtheorem{Algorithm*}{Algorithm}

\fancypagestyle{plain}{
  \fancyhf{}
  \fancyhead[C]{This is the authors version of the paper that has been accepted for publication in IEEE International Symposium on Networks, Computers and Communications (ISNCC'20),  20-22 October 2020, Montreal, Canada}     
  \fancyfoot[L]{This is a notice}

}
\usepackage{eso-pic}

\begin{document}

\AddToShipoutPictureBG*{%
  \AtPageUpperLeft{%
    \setlength\unitlength{1in}%
    \hspace*{\dimexpr0.5\paperwidth\relax}
    \makebox(0,-0.75)[c]{This is the authors version of the paper that has been accepted for publication in IEEE ISNCC'20,  20-22 October 2020, Montreal, Canada}
}}
\AddToShipoutPictureBG*{%
  \AtPageLowerLeft{%
    \setlength\unitlength{1in}%
    \hspace*{\dimexpr0.5\paperwidth\relax}
    \makebox(0,0.75)[c]{2020 IEEE International Symposium on Networks, Computers and Communications (ISNCC'20)}
}}

\title{Multi-stage Jamming Attacks Detection using Deep Learning Combined with Kernelized Support Vector Machine in 5G Cloud Radio Access Networks}

\author{\IEEEauthorblockN{Marouane Hachimi\IEEEauthorrefmark{1},
Georges Kaddoum\IEEEauthorrefmark{1},
Ghyslain Gagnon\IEEEauthorrefmark{1} and Poulmanogo Illy\IEEEauthorrefmark{1}}
\IEEEauthorblockA{
\\\IEEEauthorrefmark{1}Electrical Engineering Department, \'Ecole de technologie sup\'erieure, Montr\'eal, Canada
\vspace{3mm}
\\Email: marouane.hachimi.1@ens.etsmtl.ca}}
\providecommand{\keywords}[1]{\textbf{\textit{Index terms---}} #1}
\maketitle
\thispagestyle{empty}
\pagestyle{empty}

\begin{abstract}


In 5G networks, the Cloud Radio Access Network (C-RAN) is considered a promising future architecture in terms of minimizing energy consumption and allocating resources efficiently by providing real-time cloud infrastructures, cooperative radio, and centralized data processing. Recently, given their vulnerability to malicious attacks, the security of C-RAN networks has attracted significant attention. Among various anomaly-based intrusion detection techniques, the most promising one is the machine learning-based intrusion detection as it learns without human assistance and adjusts actions accordingly. In this direction, many solutions have been proposed, but they show either low accuracy in terms of attack classification or they offer just a single layer of attack detection. This research focuses on deploying a multi-stage machine learning-based intrusion detection (ML-IDS) in 5G C-RAN that can detect and classify four types of jamming attacks:  constant jamming, random jamming, deceptive jamming, and reactive jamming. This deployment enhances security by minimizing the false negatives in C-RAN architectures. The experimental evaluation of the proposed solution is carried out using WSN-DS (Wireless Sensor Networks DataSet), which is a dedicated wireless dataset for intrusion detection. The final classification accuracy of attacks is 94.51\% with a 7.84\% false negative rate.


\end{abstract}
\bigbreak
\keywords{Cloud Radio Access Network, Jamming attacks, Machine Learning-based Intrusion Detection System, Multilayer Perceptron, Support Vector Machine, Wireless Sensor Networks DataSet.}

\bigbreak
\section{INTRODUCTION}
\bigbreak

In recent years, with the strong growth in the number of customers, the consumption of data traffic from wireless terminals has drastically increased [1-4]. Moreover, according to a study conducted by the "Beijing Key Laboratory of Network System Architecture and Convergence, China" on mobile Internet penetration in the world from 2013 to 2019 [1], 48.8\% of the world's mobile phones had access to the Internet in 2014. This figure reached 61.2\% in 2018, with an average annual increase of 8.3\% in the number of mobile devices. In this context, due to the shortage of spectrum and bandwidth, traditional RANs are not able to meet the growing demands of mobile users. Cloud Radio  Access  Network (C-RANs) architectures present a promising solution to potentially increase the network flexibility and improve its performance, possibly overcoming the problems of traditional RANs [3]. In fact, real-time cloud infrastructures, cooperative radio, centralized data processing, and cloud radio access networks are increasingly sought by mobile operators to meet the requirements of end-users. Since IBM defined the concept of C-RAN in 2010, this technology has attracted significant attention around the world. The fifth generation of mobile technologies (5G) established C-RAN as its architecture to support their new mobile services and communications [4].

A C-RAN network, which has a mesh topology, is composed of three main components: Virtualized Base-Band Unit (BBU) pool, Remote Radio Head (RRH), and a fronthaul network connecting the RRHs to the BBU pool. RRHs collect wireless signals from all wireless devices and the fronthaul network sends them to the BBU pool. Thanks to a Digital Signal Processor (DSP) controller in the BBU pool, the C-RAN network can re-assign the fronthaul network to meet the changing traffic needs of mobile devices [5].

Recently, because of their vulnerability to malicious attacks, the security of C-RAN networks has drawn special attention and concern [6]. Due to the open nature of wireless networks, both authorized and illegitimate users can access the communication channel. Thus, C-RANs inherit all the attacks that can be performed on wireless networks, including the most popular jamming attacks. Due to their ability to easily disable radio channels that use strong high-level security measures, jamming attacks represent the most serious security threat in the wireless communication field. Several forms of this attack can be used against C-RAN, namely constant jamming, random jamming, deceptive jamming, and reactive jamming [7].

In this context, Intrusion Detection  System (IDS) have been developed to enhance the security of the network. Signature-based intrusion detection (S-IDS) is a valuable technology which could protect C-RAN networks against the known attacks. However, this method fails to identify new attacks [8]. On the other hand, anomaly-based intrusion detection (A-IDS) can resolve this limitation by detecting unknown or novel attacks. Among various anomaly-based intrusion detection techniques, machine learning-based intrusion detection (ML-IDS) shows great potential. In this direction, different solutions have been proposed to tackle security threats.

Syed \textit{et al.} [9] proposed a new radio modulation network-based intrusion detection system for jamming attacks named LIDS. They implemented two LIDS algorithms based on the Kullback Leibler Divergence (KLD) and Hamming distance (HD). The detection rates achieved are of 98\% and 88\%, respectively with a 5\% false positive rate. Imen \textit{et al.} [10] designed an intrusion detection mechanism to limit DoS attacks in Wireless Sensor Networks. They implemented five machine learning algorithms to detect and classify DoS attacks. Oscar \textit{et al.} [11] presented a machine learning-based jamming detection approach capable of detecting constant and reactive jammers under various scenarios in 802.11 networks. Yi \textit{et al.} [12] presented a machine learning method for launching jamming attacks in wireless communications and also introduced a defense strategy. For vehicular ad hoc networks (VANETs). Dimitrios \textit{et al.} [13] presented a method for detecting and clustering radio frequency (RF) jamming attacks based on the use of unsupervised machine learning. Thi \textit{et al.} [14] designed a machine learning based IDS to classify four DoS attacks in Wireless Sensor Networks (WSN).

This work aims to deploy a new multi-stage ML-IDS with a double detection check against jamming attacks which ensures a high attack detection accuracy. Precisely, our main contributions are summarized as follows:
\begin{itemize}

\item We propose a new ML-IDS concept based on supervised and deep learning for the detection and classification of jamming attacks.

\item We enhance the security of C-RAN networks by deploying a high accuracy multi-stage jamming attack detection mechanism.

\item We implement our final solution into the BBU pool without greatly affecting the latency.

\end{itemize}
\bigbreak
\section{Types Of Jammers And Intrusion Detection Systems}
\bigbreak

In this section we will go through the existing types of jammers then we will justify the chosen detection method.

\subsection{Types Of Jammers}

\vspace{1mm}

A jammer is an equipment that can disturb a node's signal by increasing its power spectral density (PSD). There are several types of jammers that may be used against C-RAN networks, namely: constant jammer, random jammer, deceptive jammer, and reactive jammer. 

\vspace{1mm}

\subsubsection{Constant jammer} 
a constant jammer continuously produces radio signals completely random. They do not follow any underlying MAC protocol and are only random bits.
\subsubsection{Random jammer}
a random jammer works randomly in two states; sleep and jamming. During the sleep state, it is idle and during the jamming state, it acts as a constant jammer.
\subsubsection{Deceptive jammer}
as deceptive jammer is a constant jammer that continuously transmits regular packets. It is more difficult to detect because it transmits legitimate packets instead of random bits.
\subsubsection{Reactive jammer}
a reactive or intelligent jammer is activated when it detects a transmission on the channel and starts sending illegitimate packets. If the channel is inactive, it stays inactive and continues sensing the channel.

\vspace{2mm}
\subsection{The Chosen Detection Method}
\bigbreak



IDSs are strategically placed on a network to detect threats and monitor network traffic. IDS uses network or host-based approaches to recognize attacks by collecting data from network systems and sources and analyzing it to identify potential threats.



As  we  can  see  in  Figure  1, there are several intrusion detection methods, the most popular being \textit{signature-based}. Signature-based detection is governed by a set of rules used to match models in network traffic. It detects well-known attacks; however, it has a major drawback since it is incapable of identifying new attacks. On the other hand, \textit{anomaly based-IDS} detect unknown or novel attacks. Anomaly-based IDS detect attacks that have not been dealt with before.

\bigbreak
\includegraphics[scale=0.22]{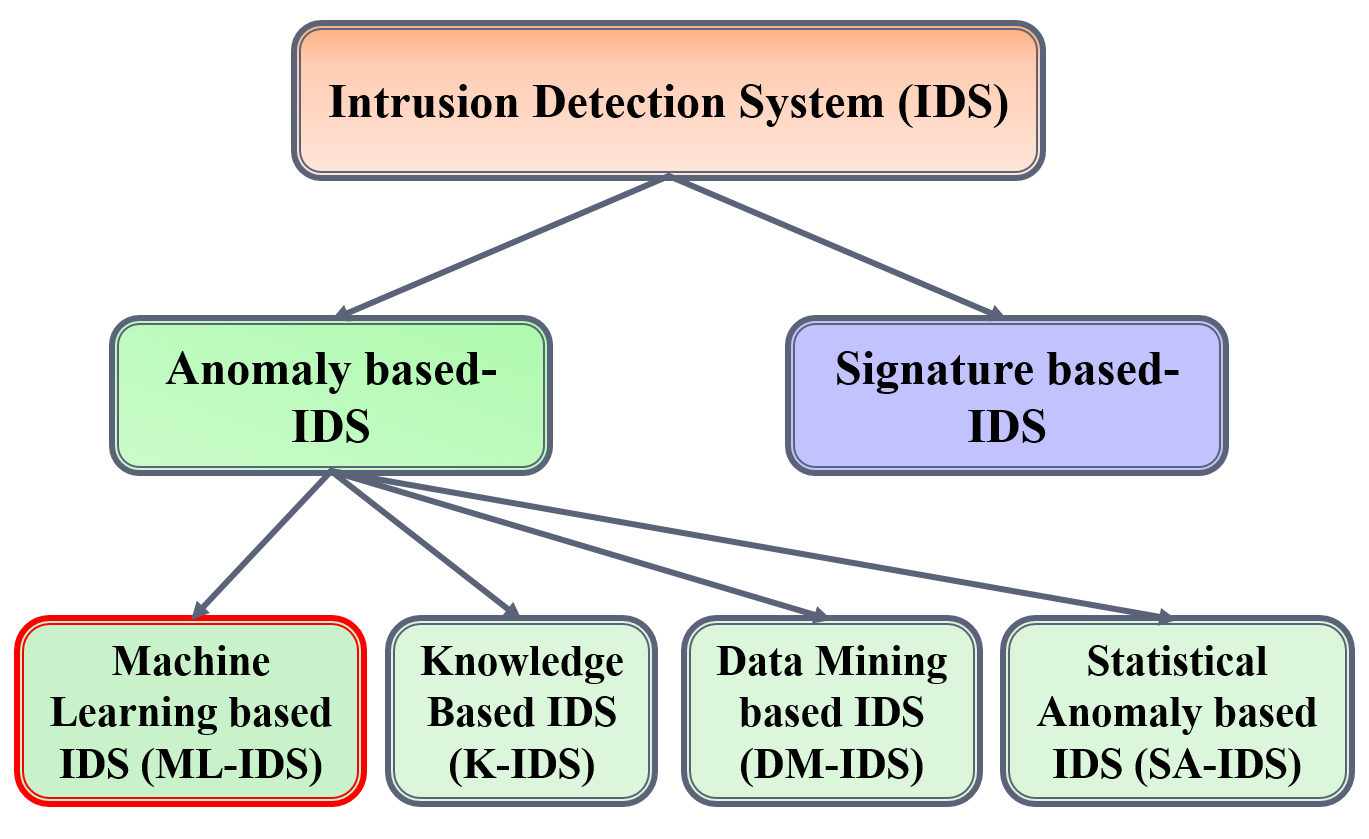} 
\begin{center}
Fig. 1: Hierarchical Classification of IDS [8].
\end{center}
\bigbreak

Among the various \textit{anomaly based-IDS} techniques such as Machine Learning based IDS (ML-IDS), Knowledge Based IDS (K-IDS), Data Mining based IDS (DM-IDS), Statistical Anomaly based IDS (SA-IDS), the most promising one is ML-IDS since it is capable of gradually improving its performance by learning over time while performing a given task.

\bigbreak
\section{The Proposed System Model}
\bigbreak

This section presents the proposed system model by first exhibiting the architecture deployed and then justifying the chosen implemented classifiers.

\subsection{The Deployment Architecture}
\vspace{2mm}
The deployed architecture of the proposed ML-IDS in C-RAN architecture is shown in Figure 2 with its mesh topology using Low Energy Aware Cluster Hierarchy (LEACH) routing protocol. LEACH is an adaptive and self-organized clustering protocol in WSNs that is characterized by its simplicity and low energy. LEACH assumes that the base station (BS) is fixed and located far from the sensor nodes.  The main idea of the LEACH protocol is to arrange nodes into clusters to distribute energy among all nodes in the network. In addition, in each cluster there is a Cluster Head (CH) node that collects the data received from the sensors in its cluster and transmits it to the BS. At the BSs we have a number of antennas distributed geographically to provide higher coverage. 

Each antenna is connected with a Remote Radio Head (RRH) through a coaxial cable, and every RRH is connected to a Base Band Unit (BBU) pool via an optical fiber which has a very low loss. The \textit{Fronthaul} is the part between the RRHs and the BBU pool which is the physical section while the part between the BBU pool and the mobile core network (internet, cloud computing resources ...) is called \textit{Backhaul} which is the virtual section of the network.

\bigbreak
\includegraphics[scale=0.175]{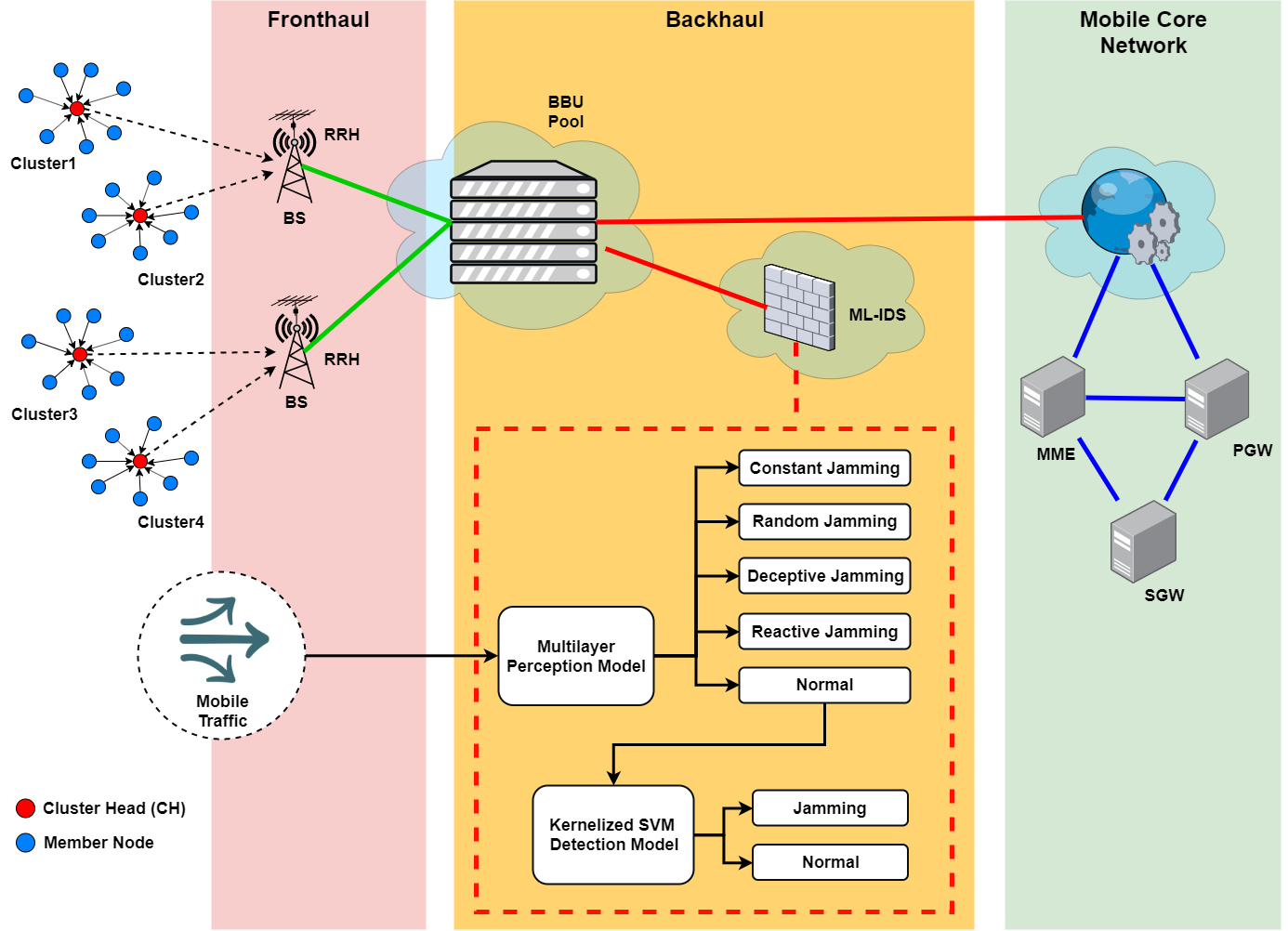} 
\begin{center}
Fig. 2: Architecture for deploying the proposed ML-IDS in C-RAN environments.
\end{center}
\bigbreak

The proposed ML-IDS collects mobile traffic flowing between the clusters and the FrontHaul, and processes it with Multilayer Perceptron (MLP). Our model classifies traffic into five classes, namely constant jamming, random jamming, deceptive jamming, reactive jamming, and normal traffic. If classified as normal by MLP, the traffic is processed again by a Kernelized Support Vector Machine (KSVM). The motivation to add a KSVM after the MLP is to reduce the false negatives that the MLP has created. In a false negative, the system decides that the situation is normal while in reality there is an attack. 

The ML-IDS is deployed into the virtualized BBU pool for several reasons. First, the virtualized BBU pool contains all the functionalities of the C-RAN network such as spectrum allocation, confidential user data, network slicing, cloud services management, etc. Thus, the BBU pool controls the entire C-RAN network. Second, the objective of a jamming attack is to add noise in the area between clusters and BSs where the BBU pool is the only part of the C-RAN that monitors this area. And finally, the BBU pool is the infrastructure that contains enough resources to run such a detection engine without greatly affecting the latency.

\subsection{The Implemented Classifiers}
\vspace{2mm}
Two classifiers have been implemented in the BBU Pool to detect the aforementioned four types of jamming attacks. Therefore, if an attack is missed with the first classifier the second one will be able to detect it. The first classifier is MLP which is a deep learning algorithm. MLP consists of a system of interconnected neurons as shown in Figure 3, which is a model representing a non-linear mapping between input and output vectors.

\begin{center}
\includegraphics[scale=0.4]{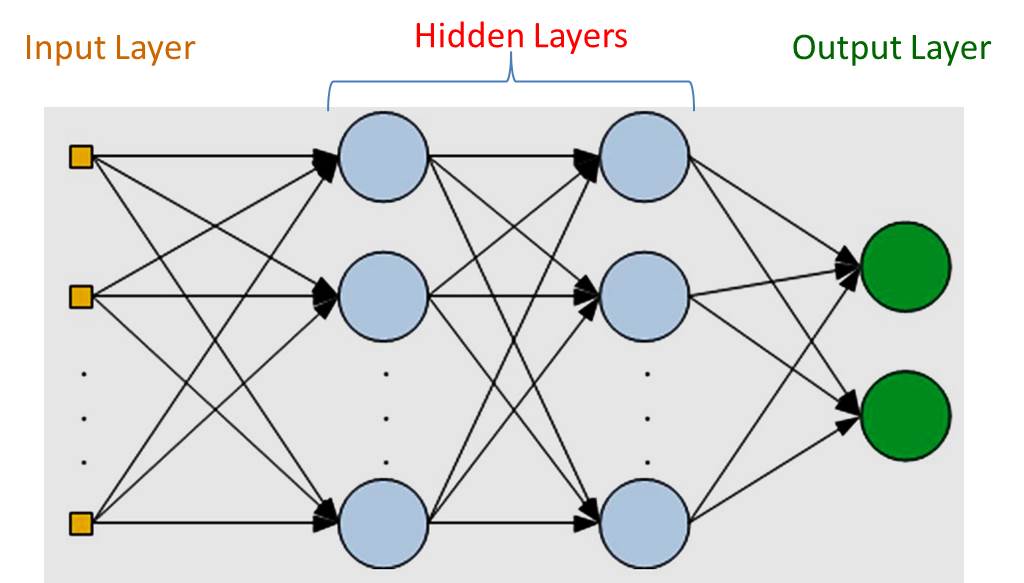}

\centering
Fig. 3: A multilayer perceptron with several hidden layers [15].
\end{center}
\bigbreak

MLP has been chosen because of the very large number of input vectors (we have 374661 vectors in our case) so the stochastic gradient drop is often the best choice (especially for classification) in terms of speed, ability, and control. MLP was chosen among all deep learning algorithms (CNN, RNN, LSTM, DBM...) because of its flexibility which allows it to be applied to different types of data. 

The second algorithm implemented is the KSVM which is an efficient binary classifier. Knowing that jamming attacks are not linearly separable in a low dimension space, the KSVM handles such situations when using a kernel function. This function maps the data in a different space where a linear hyperplane can be used to split the attacks. The process is illustrated in Figure 4. This is called the \textit{kernel trick} as the kernel function transforms the data into a higher dimensional feature space so that a linear separation is possible.

\bigbreak
\includegraphics[scale=0.2]{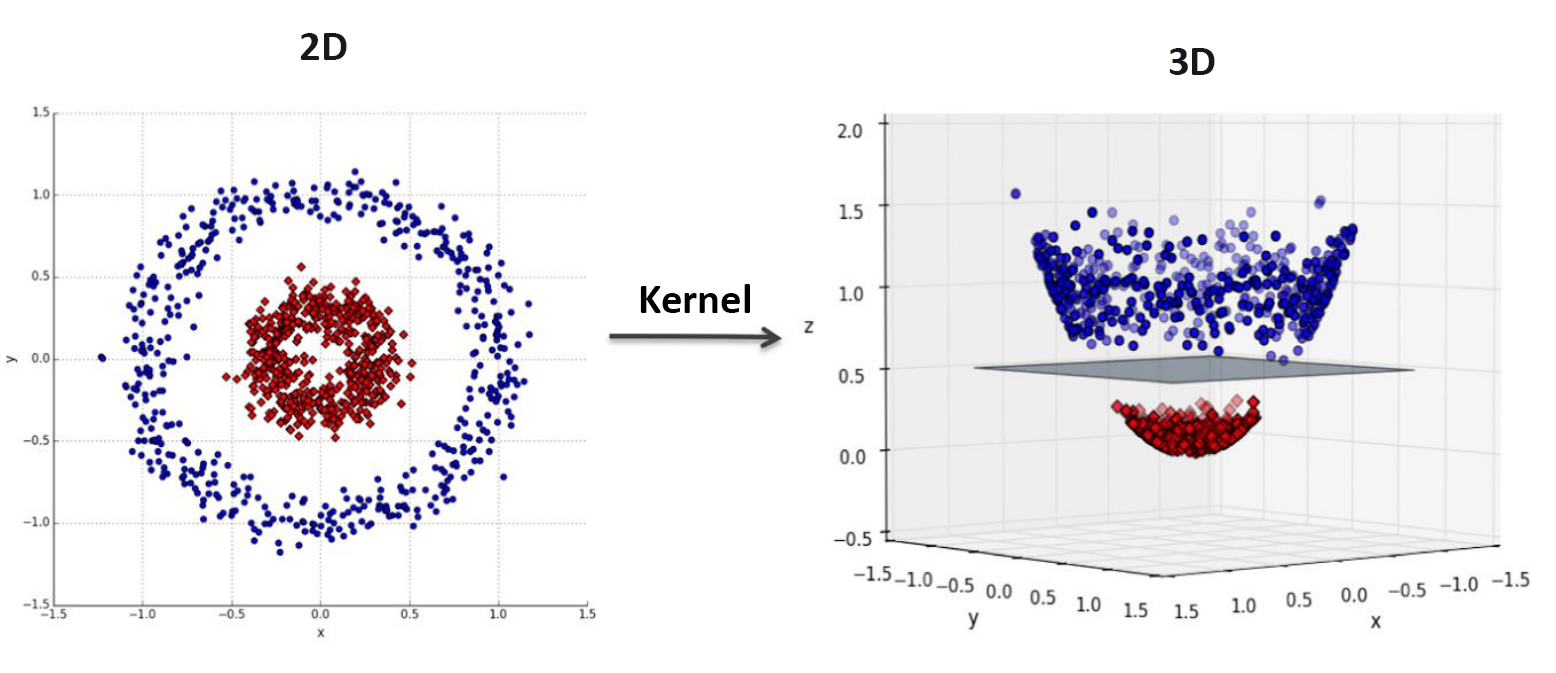} 
\begin{center}
Fig. 4: Non-linear classifier using Kernel trick [16].
\end{center}
\bigbreak

The decision limit or so-called hyperplane separating the classes has weighting coefficients given by the vector \textit{W}, which we need to estimate. The KSVM classifier tries to maximize the distance between this vector \textit{W} and the nearest points (support vectors) so that it becomes our constraint. This is equivalent to minimizing the following equation: 

\[W(\alpha)=-\sum_{i=1}^{l}\alpha_{i} + \frac{1}{2}\sum_{i=1}^{l}\sum_{j=1}^{l}y_{i}y_{j}\alpha _{i}\alpha _{j}x_{i}x_{j}\]

Subject to:
\[\sum_{i=1}^{l}y_{i}\alpha _{i}=0\]

Where \textit{l} is the number of data points in our training data, \textit{y} denote the outputs of the data points, \textit{x} is the feature vector in each training example, and $\alpha$ is the Lagrangian constant.

\bigbreak
\section{WSN-DS Description}
\bigbreak

To obtain experimental results, the WSN-DS, which is a specialized dataset for WSNs was used to classify attacks, which is a dedicated wireless dataset for intrusion detection [9]. It contains exactly 374,661 simple connection vectors, each of which includes 23 features and is labeled as normal or attack. The specific types of attacks are grouped into different categories of attacks, namely Constant jamming, Random jamming, Deceptive jamming, and Reactive jamming, in addition to the normal case (without attack).

The WSN-DS contains 23 attributes (features) to help determine the state of every node in the C-RAN network. Principal Component Analysis (PCA) is a statistical technique primarily used for dimensionality reduction which consists of selecting the attributes that contain the maximum amount of information in the order of importance, to have a model that is easier to interpret and that reduces the calculation time required. As we can see in Figure 5, we were able to extract the most important features from the 23, and these selected attributes are listed as follows:

\begin{itemize}

\item \textbf{Energy consumption}: the amount of energy consumed in the previous round.
\vspace{2mm}

At first, each node generates an arbitrary number between 0 and 1, then a threshold is computed T(n) using the formula below. If the chosen random number is less than the threshold value, the node will become a Cluster Head (CH).

\[ T(n) =
  \begin{cases}
\frac{p}{1 -p \times (r  \bmod p^{-1})} & \forall n \in N\\ 
0 & \text{ otherwise }
\end{cases}
\]
\bigbreak

where \textit{p} is the CH probability, \textit{N} is the set of nodes that have not been a CH in the last 1/\textit{p} rounds, and \textit{r} is the current round.

\item \textbf{Is CH}: A flag to distinguish whether the node is a CH (value 1) or a normal node (value 0).

\item \textbf{ADV CH send}: the number of advertise CH broadcast messages sent to the nodes.
\item \textbf{ADV SCH send}: the number of advertise TDMA schedule broadcast messages sent to the nodes.
\item \textbf{Data sent to BS}: the amount of packets of data transmitted to the RRH.
\item \textbf{Distance CH to BS}: the distance between the CH and the RRH.
\item \textbf{Data received}: the number of packets received from CHs.
\item \textbf{ADV CH receives}: the number of advertise CH messages received from CHs.
\item \textbf{Join REQ receive}: the number of join request messages received by the CHs from the nodes.
\item \textbf{Time}: the current simulation time of the node.
\end{itemize}
\bigbreak
\includegraphics[scale=0.58]{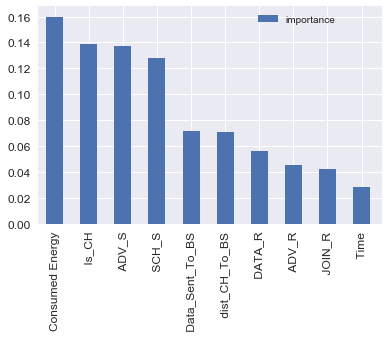} 
\begin{center}
Fig. 5: Top 10 most important features in WSN-DS.
\end{center}
\bigbreak

The WSN-DS was separated into 70\% training data and 30\% testing data. Table I shows the data separation.

\begin{table}[h]
\centering
\caption{\label{tab:table-name}Number of records used in training and testing datasets.}
\begin{tabular}{@{}ccc@{}}
\toprule
\multirow{2}{*}{\textbf{Class}} & \multicolumn{2}{c}{\textbf{Number of records used in dataset}} \\ \cmidrule(l){2-3} 
 & \multicolumn{1}{c|}{\textbf{Training set (70\%)}} & \multicolumn{1}{c|}{\textbf{Testing set (30\%)}} \\ \midrule
\multicolumn{1}{|c|}{Normal} & \multicolumn{1}{c|}{238103} & \multicolumn{1}{c|}{101963} \\ \midrule
\multicolumn{1}{|c|}{Constant jamming} & \multicolumn{1}{c|}{10233} & \multicolumn{1}{c|}{4363} \\ \midrule
\multicolumn{1}{|c|}{Random jamming} & \multicolumn{1}{c|}{6960} & \multicolumn{1}{c|}{3089} \\ \midrule
\multicolumn{1}{|c|}{Deceptive jamming} & \multicolumn{1}{c|}{4650} & \multicolumn{1}{c|}{1988} \\ \midrule
\multicolumn{1}{|c|}{Reactive jamming} & \multicolumn{1}{c|}{2316} & \multicolumn{1}{c|}{996} \\ \midrule
\textit{Sum} & \textit{262262} & \textit{112399} \\ \bottomrule
\end{tabular}
\end{table}
\bigbreak
\section{Experimentation results}
\bigbreak
The results obtained from the dataset are presented in this section. All the tasks are performed using the \textit{Python} programming language and \textit{Scikit-learn} library. Experiments were conducted on an Intel(R) Xeon(R) CPU E3-1225 v5, 16.00 GB RAM with Windows 10 Enterprise 2016 LTSB 64-bit Operating System, and x64-Based Processor. Table II show the classification accuracy obtained for each stage.



\begin{table}[h]
\caption{\label{tab:table-name}Classification accuracy for each stage.}
\centering
\begin{tabular}{@{}cc@{}}
\toprule
\textbf{Method used} & \textbf{Accuracy} \\ \midrule
\multicolumn{1}{|c|}{MLP (first stage)} & \multicolumn{1}{c|}{81,73 \%} \\  \midrule
\multicolumn{1}{|c|}{MLP + KSVM (second stage)} & \multicolumn{1}{c|}{94,51 \%} \\\bottomrule
\end{tabular}
\end{table}

To evaluate the proposed approach, We compared our multi-stage model with another work that used just MLP [10] applied to the same dataset to confirm that our approach is the most appropriate. Table III illustrates that our model can provide better accuracy of attacks classification than just the MLP model.


\begin{table}[h]
\caption{\label{tab:table-name}Global and classes accuracies.}
\begin{tabular}{@{}ccccccc@{}}
\toprule
\multirow{2}{*}{\textbf{Model}} & \multicolumn{5}{c}{\textbf{Accuracy of attacks classification}} & \multirow{2}{*}{\textbf{\begin{tabular}[c]{@{}c@{}}Global \\ Acc\end{tabular}}} \\ \cmidrule(lr){2-6}
 & \multicolumn{1}{c|}{\textit{Rando.}} & \multicolumn{1}{c|}{\textit{Const.}} & \multicolumn{1}{c|}{\textit{React.}} & \multicolumn{1}{c|}{\textit{Decept.}} & \multicolumn{1}{c|}{\textit{Norm.}} &  \\ \midrule
\multicolumn{1}{|c|}{\textbf{MLP {[}10{]}}} & \multicolumn{1}{c|}{92.8\%} & \multicolumn{1}{c|}{75.6\%} & \multicolumn{1}{c|}{99.4\%} & \multicolumn{1}{c|}{92.2\%} & \multicolumn{1}{c|}{99.8\%} & \multicolumn{1}{c|}{\textbf{91.9\%}} \\ \midrule
\multicolumn{1}{|c|}{\textbf{\begin{tabular}[c]{@{}c@{}}MLP +\\ KSVM\end{tabular}}} & \multicolumn{1}{c|}{95.3\%} & \multicolumn{1}{c|}{82.9\%} & \multicolumn{1}{c|}{99.6\%} & \multicolumn{1}{c|}{94.7\%} & \multicolumn{1}{c|}{100\%} & \multicolumn{1}{c|}{\textbf{94.5\%}} \\ \bottomrule
\end{tabular}
\end{table}

Moreover, Receiver Operating Characteristic (ROC) curve is used to visualise the performance of the classifiers. It gives us the trade-off between the True Positive Rate (TPR) and the False Positive Rate (FPR) at different classification thresholds. 

\[TPR=\frac{TP}{TP + FN}\]
\[FPR=\frac{FP}{TN + FP}\]
\bigbreak
As shown in the equations above, TPR is the proportion of observations that are correctly predicted to be positive. However, FPR is the proportion of observations that are incorrectly predicted to be positive. Figure 6 shows the ROC curve for each model. 

\bigbreak
\includegraphics[scale=0.28]{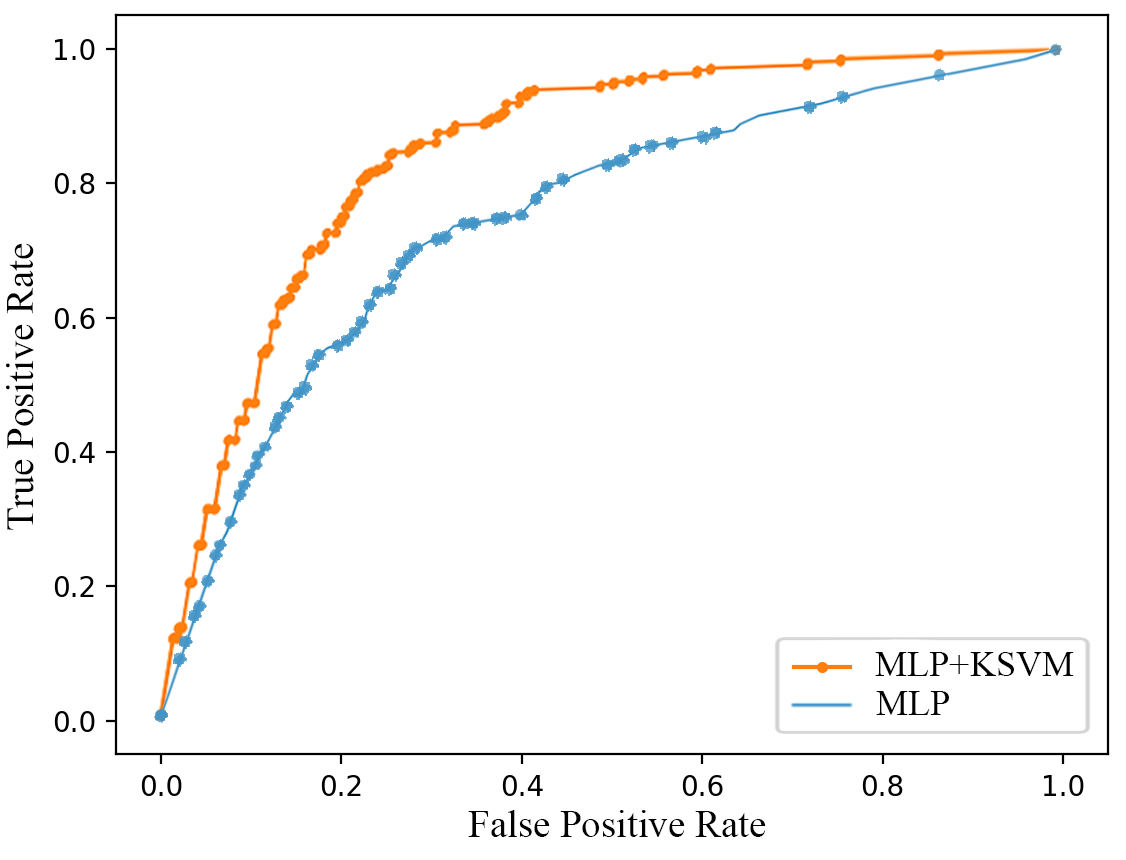} 

\begin{center}
Fig. 6: Receiver operating characteristic curve.
\end{center}
\bigbreak

These results showed that the application of MLP and KSVM to WSN-DS dataset provides a higher classification accuracy and better security in C-RAN architectures.

\vspace{1mm}
\section{Conclusion and Future Work}
\vspace{1mm}
In the present paper, we proposed an efficient multi-stage solution that can detect four different types of jamming attacks in Cloud Radio Access Networks (C-RAN) by deploying a new machine learning-based intrusion detection system (ML-IDS). We have implemented a multi-stage detection based on supervised and deep learning classifiers to reduce the number of attacks missed and to decrease the system's false negatives and false positives rates. The proposed solution guarantees a high detection and classification accuracy that can reach up to 94\%. In the future, we aim to create our wireless dataset for intrusion detection to include other types of jamming attacks such as shot noise-based intelligent jamming. In addition, several attacks that target C-RAN architectures like eavesdropping attacks, primary user emulation attacks, and impersonation attacks will be included.

\end{document}